\documentclass[12pt]{article}       
\usepackage{authblk}
\usepackage{graphicx}
\usepackage{mathptmx}      
\usepackage{amssymb}
\usepackage{amsmath}
\usepackage{color}
\usepackage{mathtools}
\allowdisplaybreaks
\usepackage{multicol}
\usepackage{multirow}
\usepackage{tikz}
\usepackage{lscape}
\usepackage{pgfplots}
\usepackage{subfigure}
\usetikzlibrary{shapes,arrows}
\usepackage{array}
\newcommand{\diff}[2]{\frac{\partial#1}{\partial#2}}
\newcommand{\dif}[2]{\frac{d\,#1}{d\,#2}}
\newcommand{\be}{\begin{equation}}
\newcommand{\ee}{\end{equation}}
\newcommand{\bal}{\begin{align}}
\newcommand{\eal}{\end{align}}

\newcommand{\ol}{\overline}

\title{A Stochastic Closure for Two-Moment Bulk  Microphysics of Warm Clouds:  Part I, Derivations}

\author{David Collins\thanks{Corresponding author, email: \texttt{davidc@uvic.ca} } ~and Boualem Khouider\thanks{Tel. 250-721-7439, email: \texttt{khouider@uvic.ca} } }
\affil{               University of Victoria,  
               PO BOX 3060 STN CSC,  \\
                Victoria, BC,  
                 Canada V8W 3P4			}

\date{8 May 2016 \\ Submitted on 2 May 2016 to \\ \textbf{Meteorology and Atmospheric Physics}}

\begin{document}

\maketitle

\begin{abstract}
We propose a methodology for stochastic parameterization of bulk warm
cloud micro-physics properties. Unlike previous bulk parameterizations,
we do not assume any particular droplet size distribution; all
parameters have physical meanings, which are recoverable from data, and
the resultant parameterization has the flexibility to utilize a variety
of collision kernels. Our strategy is a new two-fold approach to
modelling the kinetic collection equation (KCE). By partitioning the
droplet spectrum into two large bins representing cloud and rain
aggregates, we represent droplet densities as a mean plus a random
fluctuation. Moreover, we use a Taylor approximation for the collision
kernel around the centres of masses of bulk cloud and rain aggregates
which allows the resulting parameterization to be independent of the
collision kernel. Two-moment equations are thus obtained for cloud and
rain mass and number aggregates in the form of stochastic differential equations.

Based on numerical simulations of the KCE, an order of magnitude
argument on the temporal fluctuations of the evolving cloud properties
eliminates random fluctuations from the stochastic differential equations. The
physical constraints of conservation of mass and consistency of number
are used to reduce the parameterization problem to three key parameters,
controlling the strength of cloud and rain self-collection, and auto
conversion. The remaining terms form a coupled system of deterministic
ODEÕs without any ad-hoc parameters and flexible enough to accept any
collision kernel without further derivations. In the companion paper,
physical constraints are further used to constrain the parameters and
the model is validated for the case of a piece-wise polynomial kernel.

\end{abstract}

\section{Introduction}  \label{sec:intro}
Clouds are among the most poorly understood components of climate models.  Physical processes in clouds have length scales which span nine orders of magnitude \cite{HP97}.  Computers are not capable of performing the calculations necessary to resolve all of the microphysical processes that occur in a cloud that is fully contained in a model's grid cell \cite{BD12}.  Therefore, approximations and simplifications to the dynamic equations that represent these microphysical processes are necessary when these processes provide climate models with macroscopic cloud information such as the onset of precipitation and radar reflectivity \cite{CF07}.  

A sequence of three general stages result in the production of rainfall at the Earth's surface: condensation and nucleation, collision and coalescence, and precipitation \cite{WG13}.  Each of these stages is modelled as a suite of physical processes.  The processes of collision and coalescence are dominant after condensation and nucleation when water droplets are several microns in radii, and before the droplets are large enough to have a terminal velocity which exceeds typical cloud updrafts \cite{CF08}.  Bulk rate equations are a set of coupled differential equations for rain and cloud aggregates that simplify the collision and coalescence processes.

Self-collection (cloud and rain), autoconversion, and accretion are the collision and coalescence processes that comprise bulk rate equations.  Autoconversion is the only one of these processes that appears in each of the differential equations in the coupled set.  It is also the one most strongly influenced by turbulence which makes it the most difficult to accurately model \cite{RS03}.   Over the past 40+ years, more research has focused on parameterizing auto-conversion, particularly the auto conversion producing rain mixing ratio, than the other processes.  When a full suite of collision and coalescence parameterizations were derived, it is a common practice to make the remaining autoconversion processes to be diagnostic in terms of the parameterized rain auto conversion \cite{MK00,AS01,CF08}.  

Here, we propose a new methodology to derive bulk cloud microphysics equations from the kinetic coalescence equation in the case of warm clouds, but in principle it can be applied to more general situations. Our method is based on a systematic decomposition of the cloud and rain droplets into a mean and a set of random fluctuations. The latter leads to the closure of the high order moments as a sequence of Ornstein Uhlenbeck-like processes. 

For the remainder of Section \ref{sec:intro}, we briefly review the historical development of bulk parameterizations.  
In Section \ref{sec:KCE} we identify assumptions and approximations to the kinetic collection equation used in the stochastic bulk parameterization.  Section \ref{sec:SBPR} derives the four stochastic differential equations used in our parameterization.  In Section \ref{sec:MSBRP} the parameterization is closed and the parameter space is reduced to three degrees of freedom. 
Section \ref{sec:Conclu} concludes by identifying the theoretical significance of the stochastic bulk parameterization of cloud microphysical processes and its place among other parameterizations in the literature. 

The seminal work for bulk cloud microphysics parameterizations was an autoconversion parameterization as a function of cloud liquid water content $L_c$ in Kessler's 1969 paper which used a Heaviside function to terminate the autoconversion when a critical threshold was reached \cite{YL04}.  Liu and Daum (2004), hereafter referred to as LD04, improved the Kessler-Type parameterization by using a theoretical foundation to derive a dependence on cloud droplet concentration $N_c$ \cite{YL04}.  The Sunqvist-Type parameterization replaced the Heaviside function with a decaying exponential and was improved by Liu et. al. (2006), hereafter referred to as LD06,  by using a similar theoretical basis to derive a dependence on cloud droplet concentration \cite{YL06}.  Khairoutidnov and Kogan (2000), hereafter referred to as KK00,  applied the least squares method to results of many simulations of a detailed microphysics method. Their input parameters spanned a state space of liquid water content and droplet concentration $N$.  They returned a single parameterization to be used for the entire state space \cite{MK00}.  Seifert and Beheng's (2001), hereafter referred to as SB01, parameterization was derived from the kinetic collection equation.  
\begin{equation}
\begin{aligned}  
  \diff{n(x,t)}{t} = \frac12\int_{0}^x n(x-x',t)n(x',t) K(x-x',x')dx' - \\  \int_0^{\infty} n(x,t)n(x',t)K(x,x')dx' \label{eq:KCE}
\end{aligned}
\end{equation} 
where $x$ is the droplet mass: $n(x,t)$ is the number concentration, a density function: and $K(x,x')$ is the collision-coalescence kernel so that $n(x,t)n(x',t)\linebreak K(x,x')$ is the rate-density of collision-coalescence between two droplets of mass $x$ and $x'$, respectively.  As was LD04 and LD06, their's was dependent on both $L_c$ and the cloud droplet concentration $N_c$.  SB01 validated their bulk parameterization with results from a detailed microphysics method \cite{AS01}.  The analytically derived parameterizations used a specific form of the droplet size distribution (DSD)  which is not universally applicable to varying cloud types  \cite{IZ94,RW05b}.  Franklin, hereafter referred to as Fr08, applied DNS results to the parameterization given by Khairoutidnov and Kogan (2000) to get parameters as a function of the Taylor-based Reynolds number \cite{CF08} and validated the results with DNS.   Franklin's turbulent parameterizations are valid for a range of turbulent kinetic energy of 100-1500 cm$^2$s$^{-3}$.  

Kessler's work and the work of LD04 and LD06 produced only auto-conversion parameterizations for rain mixing ratio.  KK00, SB01, and Fr08 derived expressions for the auto-conversion process, and for the accretion process, as they affected rain mixing ratio. They used conservation of mass to deliver an expressions for loss of cloud mixing ratio due to these processes, and used the equation 
\begin{equation} \label{eq:xqN}
\overline x = \frac{q}{N}
\end{equation}
to deliver expressions for other terms in their parameterizations.  The auto-conversion parameterizations for the remaining quantities ($N_c$, $N_r$, $q_c$) are simply set to be functions of the rain mixing ratio.  The stochastic bulk parameterization, derived here, independently develops auto-conversion and accretion terms for each of the four evolved quantities while preserving conversation of mass and consistency of number.  These parameterizations have all used a grid-box mean value for the mixing ratio.  Wood et. al. showed that the use of a grid-box mean value for the mixing ratio results in an underprediction of the autoconversion rate \cite{RW02}.   

The use of stochastic methods in cloud microphysics is fairly new besides a couple of articles known to the authors.  Posselt et al. (2010) and Van Lier-Walqui et al. (2012) represented the uncertainty in radar reflectivity via a random process.  Krueger (1993) \cite{SK93} used a power law distribution to model the variability of entrainment plume size \cite{DP10,MvLW12}.  This was developed by Krueger et. al. (1997) to include multiple events and simulate a cumulus cloud \cite{SK97}, and further developed by Su et. al (1998) to include Kolmogorov inertial range scalings and droplet growth \cite{CS98}.  These stochastic models use probability distributions to model uncertainty (`sample-based' stochastic models).  The stochastic bulk parameterization developed herein uses stochastic differential equations (SDE) and assumes no specific probability distributions.  This method can be identified as an `SDE-based' stochastic model.  The authors are unaware of any other `SDE-based' stochastic model used to represent microphysical processes in clouds.

\section{Definitions and Assumptions} \label{sec:KCE}
For simplicity in exposition, we consider the kinetic coalescence equation (KCE) Equation \ref{eq:KCE} for warm clouds.  Large droplets that  reach certain terminal velocities escape the updraughts and fall out of the cloud as rain. The droplet spectrum is thus naturally divided into rain droplets and cloud droplets depending whether the droplet mass exceeds a heuristically chosen threshold $x^*$. In practice, $x^*$ depends on the strength of the updraughts and thus it is typically larger for deep convective clouds and smaller for shallow cumulus clouds. For convenience, we assume that $x^*$ is a fixed parameter.

Accordingly, we define the droplet numbers, $N_c,~N_r$, and mixing ratios, $q_c,~q_r$, of cloud (subscript `$c$') and rain (subscript `$r$') aggregates as follows.
\be 
N_c = \int_0^{x^*} n(x) dx, ~~ q_c=\int_0^{x^*} xn(x) dx, ~~  N_r= \int_{x^*}^{\infty} n(x) dx, ~~ q_r=\int_{x^*}^{\infty} xn(x) dx \label{eq:PM}
\ee    
We define the centre of mass of the cloud and rain aggregates as 
\be 
\bar x_c  \equiv  \frac{1}{N_c}\int_0^{x^*} xn(x) dx = \frac{q_c}{N_c}, ~~\bar x_r  \equiv  \frac{1}{N_r}\int_{x^*}^{\infty} xn(x) dx= \frac{q_r}{N_r}. \label{eq:MM}
\ee 
The framework developed below can be easily extended to more than two such bins allowing the inclusion of drizzle, for example.

\subsection{Density Approximations} \label{sec:Number}
Both number concentration density and mixing ratio density are approximated by their respective aggregate totals, normalized by the width of the spectrum on which the aggregate is defined, plus a stochastic fluctuation: 
\be  
n(x,t) = \left\{ \begin{array}{cc} \frac{N_c(t)}{x^*}+\delta_c(\omega_1;x,t)  &~0\le x<x^* \\ [0.1cm]
                                 \frac{N_r(t)}{x_m-x^*}+\delta_r(\omega_1;x,t)  &~x^*\le x<x_m. \\
                                 0  & x_m \le x  \end{array}\right. \label{eq:NC}
\ee 
\begin{equation}  
q(x,t) = \left\{ \begin{array}{cc} \frac{q_c(t)}{x^*}+\gamma_c(\omega_2;x,t) &~0\le x<x^* \\ [0.1cm]
                                 \frac{q_r(t)}{x_m-x^*}+\gamma_r(\omega_2;x,t) &~x^*\le x<x_m. \\
                                 0  &  x_m \le x \end{array}\right. \label{eq:qc}
\end{equation}
Here $x_m$ is a large enough droplet mass so that $n(x)  \approx 0$ and $q(x) \approx 0$ if $x\ge x_m$. In practice it can be taken to be the maximum possible droplet mass. The fluctuations of number concentration density $\delta_{c}(\omega_1;x,t)$ and $\delta_{r}(\omega_1;x,t)$ and mixing ratio density $\gamma_{c}(\omega_2;x,t)$ and $\gamma_{r}(\omega_2;x,t)$ are functions of random numbers $\omega_1$ and $\omega_2$, have zero mean, and are assumed to have bounded variation and prescribed covariance. Such stochastic fluctuations and corresponding random variables are further developed in Section \ref{sec:SBPR}.  Because the first approximation is used to define the four aggregate quantities in Equation $\ref{eq:PM}$ and the two densities, number concentration and mixing ratio given by Equations \ref{eq:NC} and \ref{eq:qc} respectively, are related by the physical constraint $q(x,t)=xn(x,t)$, a compatibility condition exists which is developed and resolved in $\ref{sec:CompatibilityCondition}$.

\subsection{2-D Domain: Source Droplet Pairs} \label{sec:Partitioned_Domain}
Droplets affected by collision processes typically range in radii from $\sim1$ $\mu$m to several millimetres \cite{AB98,LW07,MS02}.  Droplets smaller than this range follow the streamlines of larger droplets and grow by condensation rather than by collision and coalescence, and droplets larger than this range experience breakup and rarely exceed several millimetres \cite{HP97}.    When modelling a binary collision, the spherical water masses prior to the moment of collision are called `source droplets,' and water mass in the single post-collision droplet is called the `target droplet.'

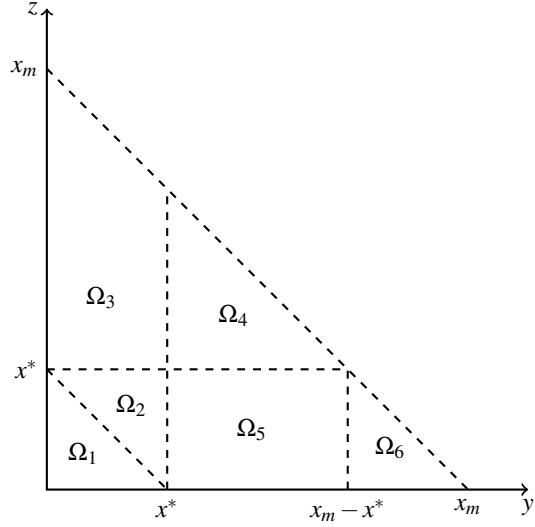
\begin{figure}
\begin{center}
\begin{tikzpicture}[thick,scale=0.80, every node/.style={scale=0.80}]
\draw [<->, thick] (0,8) --(0,0) -- (8,0);

\draw [dashed] (0,7) -- (7,0);
\draw [dashed] (0,2) -- (2,0);

\draw [dashed] (0,2) -- (5,2);
\draw [dashed] (2,0) -- (2,5);
\draw [dashed] (5,0) -- (5,2);

\node [left] at (0,7) { $x_m$};
\node [below] at (7,0) { $x_m$};
\node [left] at (0,2) { $x^*$};
\node [below] at (2,0) { $x^*$};
\node [left] at (0,8) { $z$};
\node [below] at (8,0) { $y$};
\node [below] at (5,0) { $x_m-x^*$};

\node [right] at (0.2,0.6) { $\Omega_1$ };
\node [right] at (1.0,1.4) { $\Omega_2$ };
\node [right] at (0.5,3.2) { $\Omega_3$ };
\node [right] at (2.7,2.9) { $\Omega_4$ };
\node [right] at (3.0,1) { $\Omega_5$ };
\node [right] at (5.3,0.7) { $\Omega_6$ };

\end{tikzpicture} 
\caption[Droplet Mass Aggregates]{The domain of pre-collision droplet pairs is partitioned into six regions defining the collision and coalescence processes: cloud self-collection ($\Omega_1$), auto-conversion ($\Omega_2$), accretion ($\Omega_3$, $\Omega_5$, $\Omega_6$), and rain self-collection ($\Omega_4$). \label{fig:mass_aggregates} }
\end{center} 
\end{figure}

Two orthogonal spectra of source droplets define a two dimensional domain.  The positive quadrant within this domain identifies all possible pre-collision droplet pair combinations as shown in Figure $\ref{fig:mass_aggregates}$.  The separation between rain and cloud droplets on this graph is indicated by a separation mass $x^*$.  The mass of the separation threshold $x^*$ between rain and cloud droplets and the largest droplet mass $x_m$, are indicated on the graph.  In subsequent subsections, the kinetic collection equation is integrated using $x_m$ in upper integration bounds and $x_m \rightarrow \infty$.  

The shape of the domain identified in Figure $\ref{fig:mass_aggregates}$ is produced from a change of variables:
\begin{equation} \label{eq:change_variables}
 z=x-x' ~ \text{ and } ~ y=x'
 \end{equation} 
 where the original two variables ($x-x'$ and $x'$) represent the two source droplets in the first integral in Equation $\ref{eq:KCE}$.  The change of variables is helpful to integrate the partial  moment on the rain portion of the spectrum.  The distinction between regions $\Omega_5$ and $\Omega_6$ facilitates the integration.  Both the lower and upper bounds of integration in region $\Omega_6$ contain $x_m$ which is taken in the limit to infinity.  Therefore all results in this region will be zero.  However, the calculations are detailed in the derivations for completeness.

The pre-collision droplet pair combinations are partitioned in Figure $\ref{fig:mass_aggregates}$ into four aggregates: cloud-cloud ($\Omega_1 \cup \Omega_2$), rain-cloud ($\Omega_3$), cloud-rain ($\Omega_5 \cup \Omega_6$), and rain-rain ($\Omega_4$).  The collision-coalescence processes are represented as follows: cloud self-collection ($\Omega_1$), auto-conversion ($\Omega_2$), accretion ($\Omega_3$, $\Omega_5$, $\Omega_6$), and rain self-collection ($\Omega_4$).  The three regions given by $\Omega_3$, $\Omega_4$, and $\Omega_5 \cup \Omega_6$ each span a complete aggregate and are each represented by a single collision-coalescence process.   Therefore, the integral of the fluctuations for each of these regions is zero. In Section $\ref{sec:SBPR}$, we show that the state space mean of the fluctuations is zero for a complete aggregate. 

The bounds of integration for the six regions in Figure $\ref{fig:mass_aggregates}$ are explicitly detailed as  follows:
\begin{equation}
\begin{aligned}  
& \int_{\Omega_1} d\Omega_1 \equiv  \int_0^{x^*} \int_{0}^{x^*-y} dzdy  & \text{cloud self-collection}  \\ 
& \int_{\Omega_2} d\Omega_2 \equiv  \int_0^{x^*} \int_{x^*-y}^{x^*} dzdy  & \text{auto-conversion}  \\ 
& \int_{\Omega_3} d\Omega_3 \equiv   \int_0^{x^*} \int_{x^*}^{x_m-y} dzdy & \text{accretion}  \\
&  \int_{\Omega_4} d\Omega_4 \equiv  \int_{x^*}^{x_m-x^*} \int_{x^*}^{x_m-y} dzdy & \text{rain self-collection}  \\
& \int_{\Omega_5} d\Omega_5 \equiv  \int_{x^*}^{x_m-x^*} \int_0^{x^*} dzdy  & \text{accretion}  \\ 
&  \int_{\Omega_6} d\Omega_6 \equiv  \int_{x_m-x^*}^{x_m} \int_0^{x_m-y} dzdy & \text{accretion}  \label{eq:domain_omega}
\end{aligned}
\end{equation}
  
In 2007 Wang and collaborators presented a figure (their Figure 2) constructed with pairs of source droplets.  However, they identified bin indices on a pair of discretized spectra and omitted references to the types of collision processes associated with the distinctive regions of the domain.  Figure \ref{fig:mass_aggregates} replaces the bin indices with physically relevant collision processes.

\subsection{Collision Kernel Expansion} \label{sec:collision_kernel}
A Taylor approximation of $K(x,x')$ is centred around $\bar x_c$ and $\bar x_r$ to obtain closed forms for the bulk equations to be developed in Section $\ref{sec:SBPR}$. 
\begin{equation}
\begin{aligned}  
K(x,x') = K(\bar x, \bar x') & +  \diff{K(\bar x,\bar x')}{x} (x-\bar x) + \diff{K(\bar x,\bar x')}{x'} (x'-\bar x') ~+ \\
 &\frac{\partial^2 K(\bar x,\bar x')}{2~\partial x\partial x'} (x-\bar x)  (x'-\bar x') + \frac{\partial^2 K(\bar x,\bar x')}{2~\partial x^2} (x-\bar x)^2 +  \\ & \frac{\partial^2 K(\bar x,\bar x')}{2~\partial {x'}^2} (x'-\bar x')^2 + \text{h.o.t}, \label{eq:KE}
\end{aligned}
\end{equation}

More precisely we have bins with $a\le x,\bar x < b$ and $a'\le x',\bar x'< b'$ s.t. $\bar x=\bar x_c$ ($\bar x'=\bar x_c$) if $a=0$ and $b=x^*$ ($a'=0$ and $b'=x^*$), and $\bar x=\bar x_r$ ($\bar x'=\bar x_r$) if $a=x^*$ and $b=\infty$ ($a'=x^*$ and $b'=\infty$). 

The first six terms of the Taylor approximation yield a combination of constant, linear, and product kernels considered in Drake's paper \cite{RD72} to derive benchmark exact solutions for the coalescence equation. A combination of those benchmark solutions could be used here to close the equations without any further approximation of $n(x,t)$.  However, that route increases the number of terms in the resultant bulk parameterization five-fold.  By using only the constant term in the Taylor approximation the number of terms and associated computational costs mirror that of existing bulk microphysical parameterizations.  However, unlike existing parameterizations, which are restricted to a chosen collision kernel applied during the derivation of the kinetic collection equation, the stochastic bulk parameterization can accept a wide variety of collision kernels without any further derivations.  We consider the stochastic correction terms to be in some way representative of the Taylor error terms.

\section{Stochastic Bulk Model} \label{sec:SBPR}
The density approximations given by Equations \ref{eq:NC} and \ref{eq:qc} are used when taking moments of Equation $\ref{sec:KCE}$.  By construction, the instantaneous mean, w.r.t. droplet size, of the stochastic fluctuations are zero.  The sum of the density approximations over an entire region are zero: 
\begin{equation}
N_c(t) = \int_0^{x^*} n(x,t) dx  =  \int_0^{x^*}  \frac{N_c(t)}{x^*}+\delta_c(\omega_1;x,t) dx  =  N_c(t) +  \int_0^{x^*} \delta_c(\omega_1;x,t) dx \nonumber
\end{equation}
Consequently, 
\begin{equation} \label{eq:mean0}
0 = \frac1{x^*} \int_0^{x^*} \delta_c(x) dx \equiv \left< \delta_c(x) \right>.
\end{equation}
where only the droplet mass dependency is expressed to emphasize that the mean is taken over an interval in the droplet spectrum and at an instant in time.  Similarly, we have 
\begin{equation}
\begin{aligned}  
&   \int_{x^*}^{x_m} \delta_r(x) dx \equiv \left<\delta_r(x) \right>=0, ~~~ \int_0^{x^*} \gamma_c(x) dx \equiv \left<\gamma_c(x)\right>=0, \text{ and }  \\  &  \int_{x^*}^{x_m} \gamma_r(x) dx \equiv \left<\gamma_r(x)\right>=0 \text{ for all } t. 
\end{aligned}  
\end{equation}
Moreover, we assume that $\delta_c$ and $\delta_r$ are spatially correlated with non-trivial covariance functions.  Using the change of variables from Equation \ref{eq:change_variables} for the pre-collision droplet masses, the product fluctuation terms can be expressed as:
\begin{align}
\phi_{jk} = \int_{\Omega_i} { \delta_j(\cdot) \delta_k(\cdot) } d\Omega_i,~~ \text{ or } ~~&\phi_{jk}= \int_{\Omega_i}{ \delta_j(\cdot) \gamma_k(\cdot)\ } d\Omega_i
\end{align}
where $\Omega_i$ is a domain identified in Figure $\ref{fig:mass_aggregates}$ and $\left(j,k \in \{c,r\}\right)$, and the fluctuations are a function of either $z$ or $y$ and are thus represented as, for example, $\delta_k(\cdot)$.

The stochastic terms are discussed using the change of variables in Equation \ref{eq:change_variables}.  The symmetry of $\Omega_1$, w.r.t the line $z=y$, eliminates the need to distinguish between the integration variables in the indexing scheme.  There are two stochastic processes associated with number concentration fluctuations; one w.r.t $`z$' and the other w.r.t. $`y$.'  Due to the symmetry of $\Omega_1$, they are equivalent and are combined in the second term on the rhs of Equation $\ref{eq:Ncee}$.  No further identification of fluctuations is necessary for the indexing of processes within this differential equation.  The change of variables in Equation $\ref{eq:change_variables}$ is used to define the following stochastic processes:

As the cloud properties evolve, the aggregate fluctuations and aggregate product fluctuations change.  Consider the evolution of these instantaneous statistics to be a stochastic process:
\[
X_s(t) \equiv 
\begin{cases} 
\frac{\kappa}{\Phi} \int_{\Omega_i} \delta_j(\cdot,t) d\Omega_i  \hspace{1.8cm} &\text{mean, at an instant, of a set of} \\ & \text{number concentration fluctuations}   \vspace{0.25cm}\\ 
\frac{\kappa}{\Phi} \int_{\Omega_i} \gamma_j(\cdot,t) d\Omega_i  \hspace{1.8cm} &\text{mean, at an instant, of a set of}  \\ & \text{mixing ratio fluctuations}  \vspace{0.25cm}\\ 
\frac{1}{\Phi} \phi_{jk}(t)   \hspace{0.5cm} &\text{instantaneous product fluctuation}
\end{cases} 
\]
where $s=s(i,j,k,\Phi)$ is an indexing scheme to be developed in the following subsections, and $\kappa$ is a normalization constant from Equations (\ref{eq:NC} and \ref{eq:qc}), and $\Phi$ is an appropriate function of cloud properties ($N_c, N_r, q_c, q_r$) which serves to non-dimensionalize the stochastic process.  
We assume that the stochastic processes $X_s(t)$ are time homogeneous with a fixed mean $\mu_s$ and variance $\sigma_s^2$.  We thus set
\be  \label{eq:SP}
X_s(t)  =   \mu_{s}+\sigma_{s}\xi_{s}(t)
\ee
where $\xi_{s}(t)$ is a homogeneous process with mean zero and variance one.

\subsection{Cloud Number Concentration} \label{sec:CNC}
The evolution of cloud number concentration is expressed as a partial moment of the kinetic collection equation over the cloud droplet portion of the spectrum.
\begin{equation}
\begin{aligned} 
\dif{N_c}{t} =  \frac12   \int_{0}^{x^*} \int_0^x & n(x-x') n(x') K(x-x',x')dx' dx - \\ & \int_{0}^{x^*} \int_0^{\infty} n(x) n(x') K( x, x')dx' dx.  \nonumber 
\end{aligned}
\end{equation}
The substitutions for the number concentration density are used, and kernel values at the mean mass of each partition requires splitting the second double integral into two terms:
\begin{equation}  \label{eq:Ncsecond}
\begin{aligned} 
\dif{N_c}{t} & = \frac{1}2  \int_0^{x^*} \int_0^x  \left[ \frac{N_c}{x^*}+\delta_c(x-x')\right]  \left[ \frac{N_c}{x^*}+\delta_c(x')\right] K(\ol x_c,\ol x_c) dx' dx \\
- & \int_0^{x^*} \int_0^{x^*} n(x) n(x') K(\ol x_c,\ol x_c)dx' dx - \int_0^{x^*} \int_{x^*}^{\infty} n(x) n(x') K(\ol x_c,\ol x_r)dx' dx.  
\end{aligned}
\end{equation}
Expanding the substitutions for number concentration density gives three terms in the first integral that contain stochastic fluctuations.  An abbreviated notation is used for the autoconversion/cloud self-collection and accretion kernels:  $K_{cc}=K(\bar x_c,\bar x_c)$ and $K_{cr}=K(\bar x_c,\bar x_r)$, respectively.
\begin{equation}  \label{eq:Ncexpand}
\begin{aligned}  
\dif{N_c}{t} = \frac{K_{cc}}2 &  \int_0^{x^*} \int_0^x  \frac{N_c^2}{(x^*)^2} + \frac{N_c}{x^*} \delta_c(x-x')+ \frac{N_c}{x^*} \delta_c(x')  + \delta_c(x-x')\delta_c(x') dx' dx \\ 
- &K_{cc} \int_0^{x^*} \int_0^{x^*} n(x) n(x') dx' dx - K_{cr}\int_0^{x^*} \int_{x^*}^{\infty} n(x) n(x') dx' dx. 
\end{aligned}
\end{equation}
The first term in the first integral consists of only constants and thus can be integrated.  By symmetry, the second and third terms in the first integral contain stochastic fluctuations which can be integrated analytically:
\begin{equation} 
\begin{aligned}  
& \frac{K_{cc}N_c}{2x^*}  \int_0^{x^*} \int_0^x    \delta_c(x')   dx' dx ~=~  \\  & \frac{K_{cc}N_c}{2x^*}  \int_0^{x^*} \int_0^x    \delta_c(x-x')   dx' dx ~=~ \left(\frac12-\frac{\ol{x}_c}{x^*}\right)K_{cc}N_c^2    \nonumber
\end{aligned}
\end{equation}
However, this constraint limits the usability of the two-moment bulk equations because it leads to unphysical conditions that violate first principles such as conservation of mass and consistency of number concentration. Here, we relax this constraint and assume that this term is represented through a random process, denoted below by $X_{1nf}$. This relaxation is justified, for instance, by the severe truncation applied to the kernel function in Equation \ref{eq:KE}.

The second and third integrals in Equation \ref{eq:Ncexpand} are partial moments over complete partitions of the droplet size spectrum and thus can also be integrated to obtain: $K_{cc}N_c^2$ and $K_{cr}N_cN_r$, respectively.  

The double integrals of the terms containing fluctuations and product fluctuations are aggregate fluctuations over a region at an instant.  They are non-dimensionalized and represented by the stochastic processes $X_s$.  Combining the analytically integrated terms with the stochastic process terms yields a stochastic differential equation for the evolution of cloud number concentration:  
\be
\dif{N_c}{t} = \frac14K_{cc}N^2_c+K_{cc}N_c^2X_{1nf} +\frac12K_{cc}N_c^2X_{1np} -K_{cc}N_c^2-K_{cr}N_cN_r  \label{eq:Ncee}
\ee

The number in the indices of these stochastic processes, $X_s$, identify the domain according to Figure $\ref{fig:mass_aggregates}$.  The first Latin letter, `$n$' or `$m$,' indicates number concentration or mass density, respectively.  The second Latin letter, `$f$' or `$p$,' indicates fluctuation or product fluctuation, respectively.  

We have, \bal
  X_{1nf}  \equiv & \frac{1}{x^*N_c} \int_{\Omega_1} \delta_c(\cdot) d\Omega_1 = \left( \mu_{1nf}+\sigma_{1nf}\xi_{1nf} \right),  \text{where }    \nonumber \\ 
& \frac{1}{x^*N_c} \int_{\Omega_1} \delta_c(\cdot) d\Omega_1= \frac{1}{2x^*N_c} \int_{\Omega_1} \delta_c(z) d\Omega_1 + \frac{1}{2x^*N_c} \int_{\Omega_1} \delta_c(y) d\Omega_1,  \nonumber \\
 X_{1np}  \equiv &  \frac{1}{N_c^2} \int_{\Omega_1} \delta_c(z)\delta_c(y) d\Omega_1 = \left( \mu_{1np}+\sigma_{1np}\xi_{1np} \right).  \label{eq:Ncsp}
\end{align}
As already anticipated, $\xi_{1nf}$ and $\xi_{1np}$ are homogeneous processes with mean zero and variance one, and $\mu$ and $\sigma$ are the ensemble mean and standard deviation, respectively.  
Substituting Equations $\ref{eq:Ncsp}$ 
into Equation $\ref{eq:Ncee}$, and collecting like terms gives
\begin{equation} 
\begin{aligned} 
\dif{N_c}{t} = -\left(\frac34-\frac{2\mu_{1nf}+\mu_{1np}}2  \right)& K_{cc}N^2_c-K_{cr}N_cN_r+ \\ & \frac12\left[ 2\sigma_{1nf}\xi_{1nf} + \sigma_{1np}\xi_{1np} \right] K_{cc}N_c^2. \label{eq:NceX}
\end{aligned}
\end{equation}
This is the stochastic differential equation for cloud aggregate number concentration, which contains the means of stochastic processes as physically meaningful parameters.

\subsection{Rain Number Concentration} \label{sec:RNC}
The evolution of rain number concentration is also expressed as a partial moment of the kinetic collection equation.
\begin{equation} 
\begin{aligned} 
\dif{N_r}{t} = \frac12  \int_{x^*}^{\infty} \int_0^x & n(x-x') n(x') K(x-x',x')dx' dx -  \\ & \int_{x^*}^{\infty} \int_0^{\infty} n(x) n(x') K( x, x')dx' dx  \label{eq:Nree}
\end{aligned}
\end{equation}
The convenient change of variables given in Equation $\ref{eq:change_variables}$ and the substitutions for the kernel allow the first integral to be partitioned into five integrals:
\begin{equation}
\begin{aligned}  
 \frac12 &  \int_{x^*}^{\infty} \int_0^x n(x-x') n(x') K(x-x',x')dx' dx  =  \frac{K(\bar x_c,\bar x_c)}{2}  \int_{\Omega_2}   n(y)n(z)   ~d\Omega_2 +    \\
 & \frac{K(\bar x_c,\bar x_r)}{2}  \left[\int_{\Omega_3}   +\int_{\Omega_5}    +\int_{\Omega_6} \right]   n(y)n(z)   ~d\Omega ~+~ \frac{K(\bar x_r,\bar x_r)}{2}  \int_{\Omega_4}   n(y)n(z)   ~d\Omega_4 \label{eq:Nree2}
\end{aligned}
\end{equation}
The accretion terms, $\Omega_3$, $\Omega_5$, and $\Omega_6$, integrate analytically to 1/2, 1/2, and zero, respectively.  The rain self-collection term integrates to 1/2$K(\bar x_r,\bar x_r)N_r^2$.  

The truncation of a variety of turbulent and non-turbulent kernels introduces uncertainty into an otherwise deterministic expression.  The density approximations in Equations \ref{eq:NC} and \ref{eq:qc} can restore the uncertainty to these terms.  However, data from detailed simulations show that the parameter values associated with accretion are $\sim O(10^{-19})$-$O(10^{-21})$ while those associated with rain self-collection are $\sim O(10^{-1})$-$O(10^{-2})$.  Therefore, we analytically integrate the accretion terms and use stochastic processes to represent the uncertainty associated with kernel truncation in the rain self-collection term.  Using the substitutions for number concentration, the stochastic terms contained in the first and third terms on the rhs of Equation \ref{eq:Nree2} are:
\begin{equation}      \label{eq:Nrcv}
\begin{aligned}  
& \frac{K(\bar x_c,\bar x_c) }2  \int_{\Omega_2} \left[ \frac{N_c}{x^*}+\delta_c(z)\right]\left[ \frac{N_c}{x^*}+\delta_c(y)\right] d\Omega_2, \\ 
& \frac{K(\bar x_r,\bar x_r) }2 \lim_{x_m \rightarrow \infty}  \int_{\Omega_4}  \left[ \frac{N_r}{x_m-x^*}+\delta_r(z)\right]\left[ \frac{N_r}{x_m-x^*}+\delta_r(y)\right] d\Omega_4. 
\end{aligned}
\end{equation}
The second term in Equation $\ref{eq:Nree}$ partitions into two integrals and integrates easily to yield:
\be
K_{cr}N_cN_r+K_{rr}N_r^2. \label{eq:Nrst}
\ee
Expanding the substitutions for number concentration density in Equation $\ref{eq:Nrcv}$, combining that with Equation $\ref{eq:Nrst}$, and recognizing that the accretion terms cancel (1/2+1/2-1=0) produces an equation with 4 stochastic processes:
\begin{equation}
\begin{aligned}  
\dif{N_r}{t} ~=~ & \frac{K_{cc}}2  \left( \frac12N_c^2+2N_c^2X_{2nf}+N_c^2X_{2np} \right) +  \\
&  \frac{K_{rr}}2  \lim_{x_m \rightarrow \infty}\left( \frac{(x_m-2x^*)^2}{2(x_m-x^*)^2}  N_r^2 + 2N_r^2X_{4nf}+N_c^2X_{4np} \right) - K_{rr}N_r^2 \label{eq:Nrev}
\end{aligned}
\end{equation}
where, again, $K_{cc}=K(\bar x_c,\bar x_c)$, $K_{cr}=K(\bar x_c,\bar x_r)$, and $K_{rr}=K(\bar x_r,\bar x_r)$.  
As anticipated $X_{2nf}$, $X_{2np}$, $X_{4nf}$, $X_{4np}$ are stochastic processes with their own means and variances, satisfying:
\begin{equation}
\begin{aligned}  
  & X_{2nf} \equiv \frac{1}{x^*N_c}  \int_{\Omega_2}  \delta_c(\cdot) d\Omega_2 = \left( \mu_{2nf}+\sigma_{2nf}\xi_{2nf} \right),   \\
& X_{2np} \equiv \frac{1}{N_c^2}  \int_{\Omega_2}  \delta_c(\cdot)\delta_c(\cdot) d\Omega_2 =  \left( \mu_{2np}+\sigma_{2np}\xi_{2np} \right),   \\ 
& X_{4nf} \equiv \lim_{x_m \rightarrow \infty} \frac{1}{(x_m-x^*)N_r} \int_{\Omega_4}  \delta_r(\cdot) d\Omega_4 =  \left( \mu_{4nf}+\sigma_{4nf}\xi_{4nf} \right),   \\
 & X_{4np} \equiv \lim_{x_m \rightarrow \infty}  \frac{1}{N_r^2 } \int_{\Omega_4}  \delta_r(\cdot)\delta_r(\cdot) d\Omega_4 = \left( \mu_{4np}+\sigma_{4np}\xi_{4np} \right),   \\
\end{aligned}
\end{equation}
where $\xi_{2nf}$, $\xi_{2np}$, and $\xi_{4np}$ are homogeneous processes with mean zero and variance one.  When $x_m \rightarrow \infty$, we obtain  $\ol{X_{4nf}}=0$.  Taking the limit in Equation $\ref{eq:Nrev}$, substituting the three remaining means and variances, and collecting like terms gives the stochastic parameterization for rain number concentration:
\begin{equation}
\begin{aligned}  
\dif{N_r}{t} =  ~ & \left( \frac{1}4+  \frac{2\mu_{2nf}+\mu_{2np}}2 \right)  K_{cc}N_c^2  -   \left(\frac34- \mu_{4np} \right) K_{rr}N_r^2 +   \\
& \frac12( 2\sigma_{2nf}\xi_{2nf}+\sigma_{2np}\xi_{2np}) K_{cc}N_c^2 + \sigma_{4np}\xi_{4np} K_{rr}N_r^2   \label{eq:NreX}
\end{aligned}
\end{equation}
This is the differential equation for rain aggregate number concentration, which contains the means of stochastic processes as physically meaningful parameters.  

We note that the analytic and stochastic rain collection terms, 0.5$K_{rr}N_r^2$ and $(0.75- \mu_{4np}) K_{rr}N_r^2$ respectively, are equal when $\mu_{4np}=0.25$.  Results from detailed simulations produce values of $\mu_{4np}$ near 0.25 that vary depending on the kernel and the age of the cloud, as will be reported in Part II and a future publication.

\subsection{Cloud Mixing Ratio} \label{sec:CM}
The evolution of cloud mixing ratio is expressed as a partial moment of the kinetic collection equation over the cloud droplet portion of the spectrum.
\begin{equation}
\begin{aligned}  
\dif{q_c}{t} = \frac12  \int_{0}^{x^*} \int_0^x & xn(x-x') n(x') K(x-x',x')dx' dx - \\ & \int_{0}^{x^*} \int_0^{\infty} xn(x) n(x') K( x, x')dx' dx  \nonumber 
\end{aligned}
\end{equation}
Using the substitutions for the number concentration density, mixing ratio, and the collision kernel, and applying the change of variables gives
\begin{equation}
\begin{aligned}  
\dif{q_c}{t} = & \frac{1}2  \int_0^{x^*} \int_0^{x^*-y} \left[ \frac{q_c}{x^*}+\gamma_c(z)\right]\left[ \frac{N_c}{x^*}+\delta_c(y)\right] K(\ol x_c,\ol x_c) dz dy   \\
+ &  \frac{1}2  \int_0^{x^*} \int_0^{x^*-y} \left[ \frac{N_c}{x^*}+\delta_c(z)\right]\left[ \frac{q_c}{x^*}+\gamma_c(y)\right] K(\ol x_c,\ol x_c) dz dy    \\
- & \int_0^{x^*} \int_0^{x^*} xn(x) n(x') K(\ol x_c,\ol x_c)dx' dx - \int_0^{x^*} \int_{x^*}^{\infty} xn(x) n(x') K(\ol x_c,\ol x_r)dx' dx
\end{aligned}
\end{equation}
Expanding the substitutions for number concentration density and mixing ratio density produces three terms that contain stochastic fluctuations in each of the first two integrals.  The abbreviated notation for the kernels are used: $K_{cc}=K(\bar x_c,\bar x_c)$ and $K_{cr}=K(\bar x_c,\bar x_r)$.   
\begin{equation}
\begin{aligned}  
\dif{q_c}{t} = & \frac{K_{cc}}2  \int_0^{x^*} \int_0^{x^*-y}  \frac{q_cN_c}{(x^*)^2} + \frac{q_c}{x^*}\delta_c(y)  + \frac{N_c}{x^*}\gamma_c(z) + \gamma_c(z) \delta_c(y) ~ dz dy   \\
+ &  \frac{K_{cc}}2  \int_0^{x^*} \int_0^{x^*-y}  \frac{q_cN_c}{(x^*)^2} + \frac{q_c}{x^*}\delta_c(z)  + \frac{N_c}{x^*}\gamma_c(y) + \gamma_c(y) \delta_c(z) ~  dz dy    \\
- & K_{cc} \int_0^{x^*} \int_0^{x^*} xn(x) n(x') dx' dx - K_{cr} \int_0^{x^*} \int_{x^*}^{\infty} xn(x) n(x') dx' dx
\end{aligned}
\end{equation}
The first term in each of the first two integrals contain only constants and thus can be integrated.  The second term in these integrals are the same, except for a different constant, as the ones exposed in the derivation of cloud number concentration in Section \ref{sec:CNC}.  They are also represented here as the drift component of an Ornstein-Uhlenbeck-like process.  The third terms contain $\gamma_c$, which produce third moments, and yield a closure problem.
 \be \label{eq:third_moment}
  \frac{1}{x^*q_c} \int_{\Omega_1} \gamma_c(z) d\Omega_1 = \frac12 - \frac1{q_cx^*}\int_0^{x^*} z^2 n(z) dz
\ee
The third and fourth integrals are partial moments over complete regions and thus can be integrated.  Combining the analytically integrated terms with stochastic process terms yields a stochastic differential equation for the evolution of cloud number concentration:
\begin{equation}
\begin{aligned}  
\dif{q_c}{t} = \frac12 K_{cc}q_cN_c +  & K_{cc}q_cN_cX_{1nf} +   K_{cc}q_cN_cX_{1mf}+  K_{cc}q_cN_cX_{1mp}  \\ - & K_{cc}q_cN_c-K_{cr}q_cN_r  \label{eq:qcee}
\end{aligned}
\end{equation}
where $X_{1nf}$ is detailed in Section \ref{sec:CNC}, and
\begin{equation}
\begin{aligned}  
& X_{1mf} \equiv \frac{1}{x^*q_c} \int_{\Omega_1} \gamma_c(\cdot) d\Omega_1 =  \left( \mu_{1mf}+\sigma_{1mf}\xi_{1mf} \right),   \\
& X_{1mp} \equiv  \frac{1}{q_cN_c} \int_{\Omega_1} \delta_c(\cdot)\gamma_c(\cdot) d\Omega_1 =  \left( \mu_{1mp}+\sigma_{1mp}\xi_{1mp} \right),  \label{eq:qcsp}
\end{aligned}
\end{equation}
with as usual $\xi_{1mf}$ and $\xi_{1mp}$ being homogeneous processes with mean zero and variance one.  The presence of $\gamma$ in $X_{1mp}$ yields another third moment, the closure of which is handled similarly to $X_{1mf}$.

The change of variables given in Equation $\ref{eq:change_variables}$ produces two of each fluctuation integrals in region $\Omega_1$ which are summed to produce the two stochastic processes in Equation $\ref{eq:qcsp}$.  This is done in a similar fashion as $X_{1nf}$ in Equation \ref{eq:Ncsp}.  The symmetry in region $\Omega_1$ means that the duplicate processes are equivalent; thus, the coefficient in each positive term on the rhs of Equation $\ref{eq:qcee}$ is twice the coefficient in the corresponding term in Equation $\ref{eq:Ncee}$.  The sum of the second and third terms in Equation $\ref{eq:qcee}$ correspond to the second term in Equation $\ref{eq:Ncee}$ because they are all `fluctuation' terms.  
Substituting Equation $\ref{eq:qcsp}$  in to Equation $\ref{eq:qcee}$ and collecting like terms gives
\begin{equation}
\begin{aligned}  
\dif{q_c}{t} = - & \left(\frac12-  \left( \mu_{1nf}+ \mu_{1mf}+ \mu_{1mp} \right) \right)  K_{cc}q_cN_c - K_{cr}q_cN_r+  \\
& \frac12\left[  \sigma_{1nf}\xi_{1nf}  + \sigma_{1mf}\xi_{1mf} + \sigma_{1mp}\xi_{1mp} \right] K_{cc}q_cN_c \label{eq:qceX}
\end{aligned}
\end{equation}
This is the differential equation for cloud aggregate mixing ratio, which contains the means of stochastic processes as physically meaningful parameters.  

\subsection{Rain Mixing Ratio} \label{sec:RM}
The evolution of rain mixing ratio is also expressed as a partial moment of the kinetic collection equation however collisions with cloud droplets are included.
\begin{equation}
\begin{aligned}  
\dif{q_r}{t} = \frac12  \int_{x^*}^{\infty} \int_0^x & xn(x-x') n(x') K(x-x',x')dx' dx  \\ & - \int_{x^*}^{\infty} \int_0^{\infty} xn(x) n(x') K( x, x')dx' dx  \label{eq:qree}
\end{aligned}
\end{equation}
The convenient change of variables given in Equation $\ref{eq:change_variables}$ and the substitutions for the kernel allow the first integral to be partitioned into five integrals:
\begin{equation}
\begin{aligned}  
 \frac12  \int_{x^*}^{\infty} \int_0^x & n(x-x') n(x') K(x-x',x')dx' dx ~=~  \\ & \frac{K(\bar x_c,\bar x_c)}{2}  \int_{\Omega_2}   q(y)n(z)+n(y)q(z)   ~d\Omega_2 ~+~ \\ & 
  \frac{K(\bar x_c,\bar x_r)}{2}  \left[\int_{\Omega_3}   +\int_{\Omega_5}    +\int_{\Omega_6} \right]   q(y)n(z)   ~d\Omega ~+~ \\ & \frac{K(\bar x_c,\bar x_r)}{2}  \left[\int_{\Omega_3}   +\int_{\Omega_5}    +\int_{\Omega_6} \right]   n(y)q(z)   ~d\Omega ~+~  \\ 
&  \frac{K(\bar x_r,\bar x_r)}{2}  \int_{\Omega_4}   q(y)n(z)+n(y)q(z)   ~d\Omega_4
\end{aligned}
\end{equation}
The coefficients in each set of accretion terms, corresponding to $\Omega_3$, $\Omega_5$, and $\Omega_6$, integrate analytically to 1/2, 1/2, and zero, respectively.  The rain self-collection term integrates to $K(\bar x_r,\bar x_r)N_r^2$.  Using the substitutions for number concentration and mixing ratio, the remaining terms contain stochastic fluctuations:
\begin{equation}
\begin{aligned}
& \frac{K(\bar x_c,\bar x_c)}2  \int_{\Omega_2} \left( \left[ \frac{q_c}{x^*}+\gamma_c(z)\right]\left[ \frac{N_c}{x^*}+\delta_c(y)\right] + \left[ \frac{q_c}{x^*}+\gamma_c(y)\right]\left[ \frac{N_c}{x^*}+\delta_c(z)\right] \right)  d\Omega_2   \label{eq:qrcv}
\end{aligned}
\end{equation}
The second term in Equation $\ref{eq:qree}$ partitions into two integrals and integrates easily to yield:
\be
K_{cr}q_rN_c+K_{rr}q_rN_r \label{eq:qrst}
\ee
which conveniently cancel with the $q_rN_c$ accretion terms and the rain self-collection term.

The substitutions for number concentration density and mixing ratio density are expanded.  The accretion terms containing $q_rN_c$ cancel: (1/2+1/2-1=0) while the accretion terms with $q_cN_r$ sum to unity.  Combining Equations ($\ref{eq:qrcv}$ \& $\ref{eq:qrst}$) produces an equation with 2 new stochastic processes:
\begin{equation}
\begin{aligned}
\dif{q_r}{t} = & \frac{K_{cc}}{2} \left( q_cN_c+2q_cN_cX_{2nf}+2N_cq_cX_{2mf}+2q_cN_cX_{2mp} \right) +   K_{cr}q_cN_r  \label{eq:qrev}
\end{aligned}
\end{equation}
where $X_{2nf}$ is detailed in Section \ref{sec:CM}, and the remaining means and variances of the stochastic processes in Equation $\ref{eq:qrev}$ are defined as:
\bal
& X_{2mf} \equiv \frac{1}{x^*q_c} \int_{\Omega_2}  \gamma_c(\cdot) d\Omega_2 = \left( \mu_{2mf}+\sigma_{2mf}\xi_{2mf} \right),  \nonumber \\ 
& X_{2mp} \equiv  \frac{1}{q_cN_c}  \int_{\Omega_2} \gamma_c(\cdot) \delta_c(\cdot) d\Omega_2 =  \left( \mu_{2mp}+\sigma_{2mp}\xi_{2mp} \right),  \nonumber 
\end{align}
and $\xi_{2mf}$ and $\xi_{2mp}$ are homogeneous processes with mean zero and variance one.  The third moment that arises in $X_{2mf}$ is the additive inverse of the expression for the third moment in $X_{1nf}$ an shown in the rhs of Equation \ref{eq:third_moment}, and similarly $X_{2mp}$ contains a third moment which is closed by use of the stochastic process.  Taking the limit in Equation \ref{eq:qrev}, substituting the remaining means and variances and collecting like terms gives the stochastic parameterization for rain number concentration:
\begin{equation}
\begin{aligned}
\dif{q_r}{t} = & ~ \left(\frac12+ \mu_{2nf} + \mu_{2mf} + \mu_{2mp} \right) K_{cc}q_cN_c  +  K_{cr}q_cN_r   +  \\
& 2( \sigma_{2cnf}\xi_{2cnf}+  \sigma_{2cmf}\xi_{2cmf} + \sigma_{2mp}\xi_{2mp}) K_{cc}q_cN_c     \label{eq:qreX}
\end{aligned}
\end{equation}
This is the differential equation for rain aggregate mixing ratio, which contains the means of stochastic processes as physically meaningful parameters.  

\section{Numerical Simulations and Mean Stochastic Equations} \label{sec:MSBRP}
Equation 17,23, 29, 35 form a system of stochastic differential equations which approximate the evolutions of the two moments of mass and number concentration of cloud and rain aggregates. They depend on a rather  large number of parameters represented by the means and standard deviations of OU processes. The parameter values can in principle be learned from data. As a first filtration, here we use detailed simulations based on the KCE (1) to infer the bulk behavior of these parameters. This exercise will in particular allow us to eliminate the standard deviations from all the OU processes thus resulting in a set of deterministic equations in agreement with existing bulk cloud microphysics models \cite{MK00,AS01,CF08}.  Numerical simulations were performed using Bott's Linear Flux Method (LFM) \cite{AB98}.  The LFM is a popular and computationally efficient method of discretizing the kinetic collection equation. It simulates collisions of all possible two-droplet combinations at each time step, and evolves the mass density on a equidistant log-radius spectrum.  Mass doubles every four bins (130 bins) and the radius of the smallest droplet represented on the spectrum we used is 0.589 $\mu$m.  The time step used is 1 second.

The detailed simulations are used in two independent ways to eliminate the fluctuation terms from the coupled set of ordinary differential equations shown in Equations ($\ref{eq:NceX}$, $\ref{eq:NreX}$, $\ref{eq:qceX}$, $\ref{eq:qreX}$).  First, with respect to the representation of the stochastic processes as the sum of a mean and the product of a standard deviation and a homogeneous fluctuation, we show that the standard deviation term is at least one order of magnitude smaller than the mean term.  This is shown using results from detailed simulations using the KCE with four different collision kernels.  Thus we justify eliminating the fluctuation terms  which contain $\xi$ from Equations ($\ref{eq:NceX}$, $\ref{eq:NreX}$, $\ref{eq:qceX}$, $\ref{eq:qreX}$).  Second, when taking the mean of these four equations, second order fluctuation terms such as  $\ol{\delta N_c\delta N_c}$ appear.  We use the same detailed simulations to show that $\delta{N_c}$, ~ $ \delta{N_r}$, and $\delta{q_c}$, are at least three orders of magnitude smaller than $ \ol{N_c}$, ~ $\ol{N_r}$, and $\ol{q_c}$, respectively.  The rain mixing ratio, $q_r$, does not appear in the stochastic bulk equations.  

\subsection{Temporal Mean of the Stochastic Parameters} \label{sec:TMSP}
The kinetic collection equation (Eq.  \ref{eq:KCE}) is integrated for one hour and statistics are collected for the nine stochastic parameters contained in Equations ($\ref{eq:NceX}$, $\ref{eq:NreX}$, $\ref{eq:qceX}$, $\ref{eq:qreX}$),  defined as the sum of a mean and the product of a standard deviation and a homogeneous stochastic process: $X_s\equiv  \mu_s + \sigma_s \xi_s$ where $\mu_s$ and $\sigma_s$ are collected during the detailed simulations.  The results are reported in Table 1.  The  strength of the mean term relative to the standard deviation term is shown by the ratio $\sigma$/$\mu$.  The columns in Table \ref{tab:ST_param} that are headed by this ratio show that the standard deviation term is at least one order of magnitude smaller than the mean term.  These simulations were done for four kernels:  (i) the piece-wise polynomial kernel used by Seifert and Beheng (2001), (ii) a hydrodynamic kernel with collection efficiencies tabulated by Hall (1980), (iii) a turbulent kernel parameterized by Franklin (2007), and (iv) a turbulent kernel with tabulated values by Pinsky and Khain (2008).  Based on these results the contributions of randomness parts  $\xi_s$ of the processes $X_s$ can be eliminated.

\begin{landscape}
\begin{table}[t]
\begin{center}
\begin{tabular}{ c >{\centering\arraybackslash}m{1.35cm} >{\centering\arraybackslash}m{1.35cm} >{\centering\arraybackslash}m{1.35cm} >{\centering\arraybackslash}m{1.35cm} >{\centering\arraybackslash}m{1.35cm} >{\centering\arraybackslash}m{1.35cm} >{\centering\arraybackslash}m{1.35cm} >{\centering\arraybackslash}m{1.35cm}   }
\hline
 & \multicolumn{2}{c}{Polynomial} & \multicolumn{2}{c}{Hydrodynamic} & \multicolumn{2}{c}{Turbulent A} & \multicolumn{2}{c}{Turbulent B}  \\ 
\hline
Process & $\mu$ & $\sigma$/$\mu$ & $\mu$ &  $\sigma$/$\mu$ & $\mu$ &  $\sigma$/$\mu$ & $\mu$ &  $\sigma$/$\mu$  \\ 
\hline \hline
$X_{1nf}$ &  -1.3e-01  &  -1.2e-04  &  -1.9e-01  &  -1.3e-01  &   -1.4e-01 &   -3.8e-02  & -3.2e-01  &  -1.1e-01  \\

$X_{1np}$ &  -1.5e-02  &  -7.7e-06  &  -8.2e-03  &  -3.5e-04  &  -1.2e-02 &  -5.2e-05 &  1.4e-03  &  6.9e-02  \\

$X_{2nf}$ &  3.5e-02  &  8.5e-05  &  1.1e-02  &  4.9e-03  &  1.3e-02 &  1.6e-03  &  1.2e-02  &  7.0e-03  \\

$X_{2np}$ &  1.5e-02  &  1.1e-05  &  8.0e-03  &  3.9e-04  &  1.2e-02 &  5.2e-05 &  3.4e-03  &  5.2e-03  \\

$X_{4nc}$ & 1.8e-01  &  2.3e-01  &  2.8e-01  &  3.4e-01  &  1.8e-01 & 3.1e-01 &  1.6e-01  &  4.5e-02  \\

$X_{1mf}$ &  -3.8e-01  &  -3.8e-03  &   -4.8e-01  &  -2.1e-02  &  -3.6e-01 &  -5.7e-03  &  -7.1e-01  &  -1.4e-02  \\

$X_{1mp}$ &  -7.7e-03  &  -2.8e-03  &  -7.3e-04  &  -4.9e-02  &  -5.2e-03 &  -1.3e-03  &  2.8e-02  &  5.2e-03  \\

$X_{2mf}$ &  -3.0e-01  &  -8.5e-05  &  -3.4e-01  &  -6.7e-04  &  -3.3e-01 &  -4.2e-04  &  -2.9e-01  &  -1.2e-03  \\

$X_{2mp}$ &  7.6e-03  &  2.9e-03  &  7.8e-04  &  4.3e-02  &  5.2e-03 & 1.3e-03  &  -6.8e-03  &  -2.6e-02  \\
\hline
\end{tabular} 
\caption[Stochastic Parameters]{The first column lists the nine stochastic parameters.  The results from four kernels are given.  The Polynomial kernel uses Long's sum of squared masses for cloud droplets and Golovin's   linear sum of masses for rain droplets as presented in Seifert and Beheng (2010). \nocite{AS01} The hydrodynamic kernel uses collision efficiencies given by Hall (1980). \nocite{WH80} Turbulent A is Franklin's (2007) parameterized turbulent kernel, and Turbulent B is a tabulated kernel by Pinsky and Khain (2008). \nocite{MP08}  \label{tab:ST_param} }
\end{center}
\end{table}
\end{landscape}


\subsection{Temporal Mean of the Evolving Quantities} \label{sec:TMEQ}
As a confirmation of disregarding the fluctuation terms, we take the mean of these four stochastic equations ($\ref{eq:NceX}$, $\ref{eq:NreX}$, $\ref{eq:qceX}$, $\ref{eq:qreX}$) and show that the fluctuation terms vanish.  We substitute a mean and a fluctuation for the evolved quantities:  $N_c=\ol{N_c}+\delta N_c$ where $\ol{N_c}$ is the mean.  The mean $\ol{N_c}$ and $\delta N_c$ are computed from the detailed simulations during the complete evolution.   Taking the means of each evolved quantity ($N_c$, $N_r$, $q_c$, $q_r$) produces terms containing factors such as $\ol{\delta N_c\delta N_c}$ which are not necessarily zero.  We use detailed simulation results above to show that these terms are orders of magnitude smaller than the deterministic terms.  Fluctuations of the evolving quantities at each time step are collected.  After the evolution is complete, the fluctuations for each quantity are averaged, and each instantaneous fluctuation is compared with its respective temporal mean.  

\begin{figure}[thb!]
\centering{
	\includegraphics[width=.45\textwidth,trim=30 185 35 175,clip]{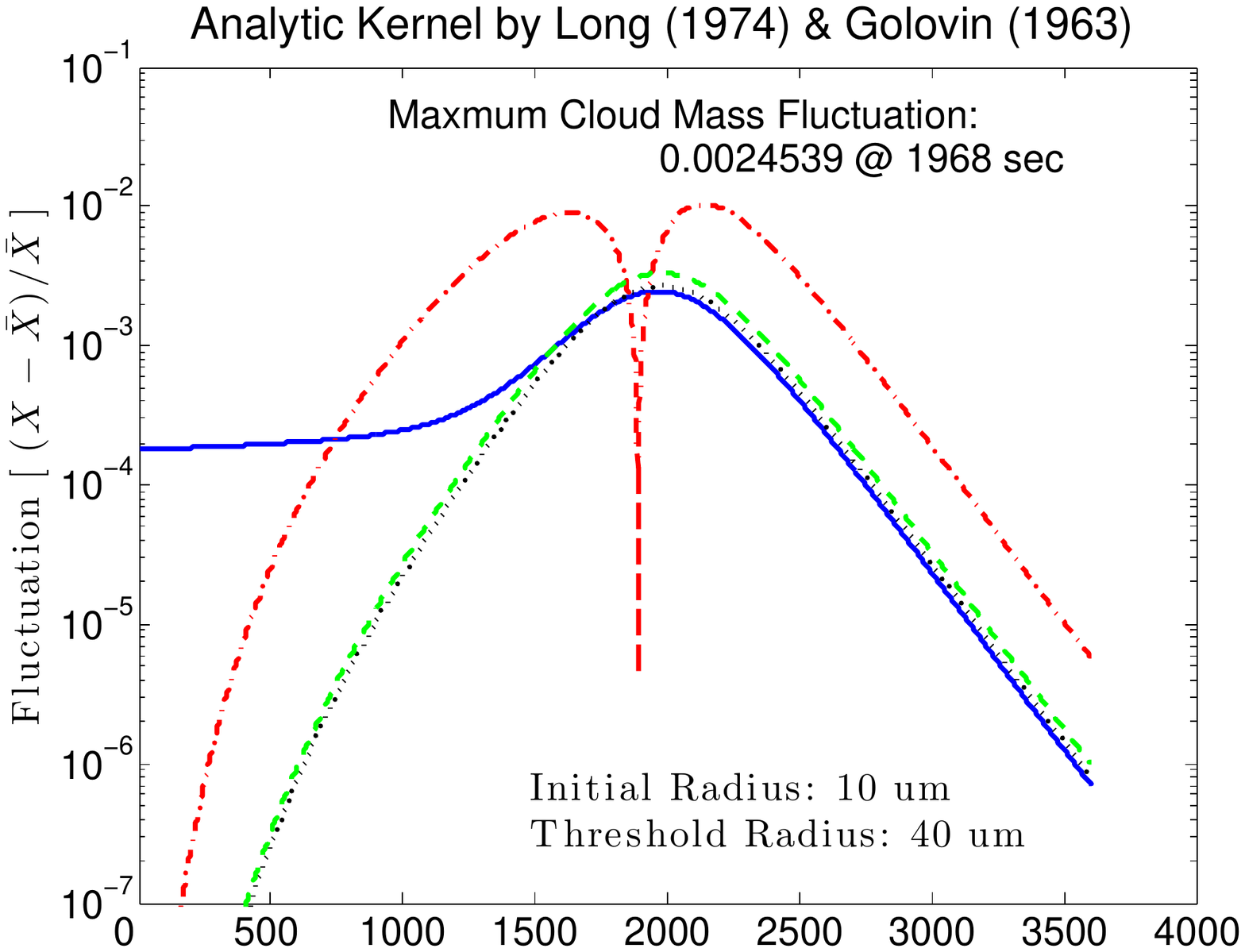} 
	\includegraphics[width=.45\textwidth,trim=30 185 35 175,clip]{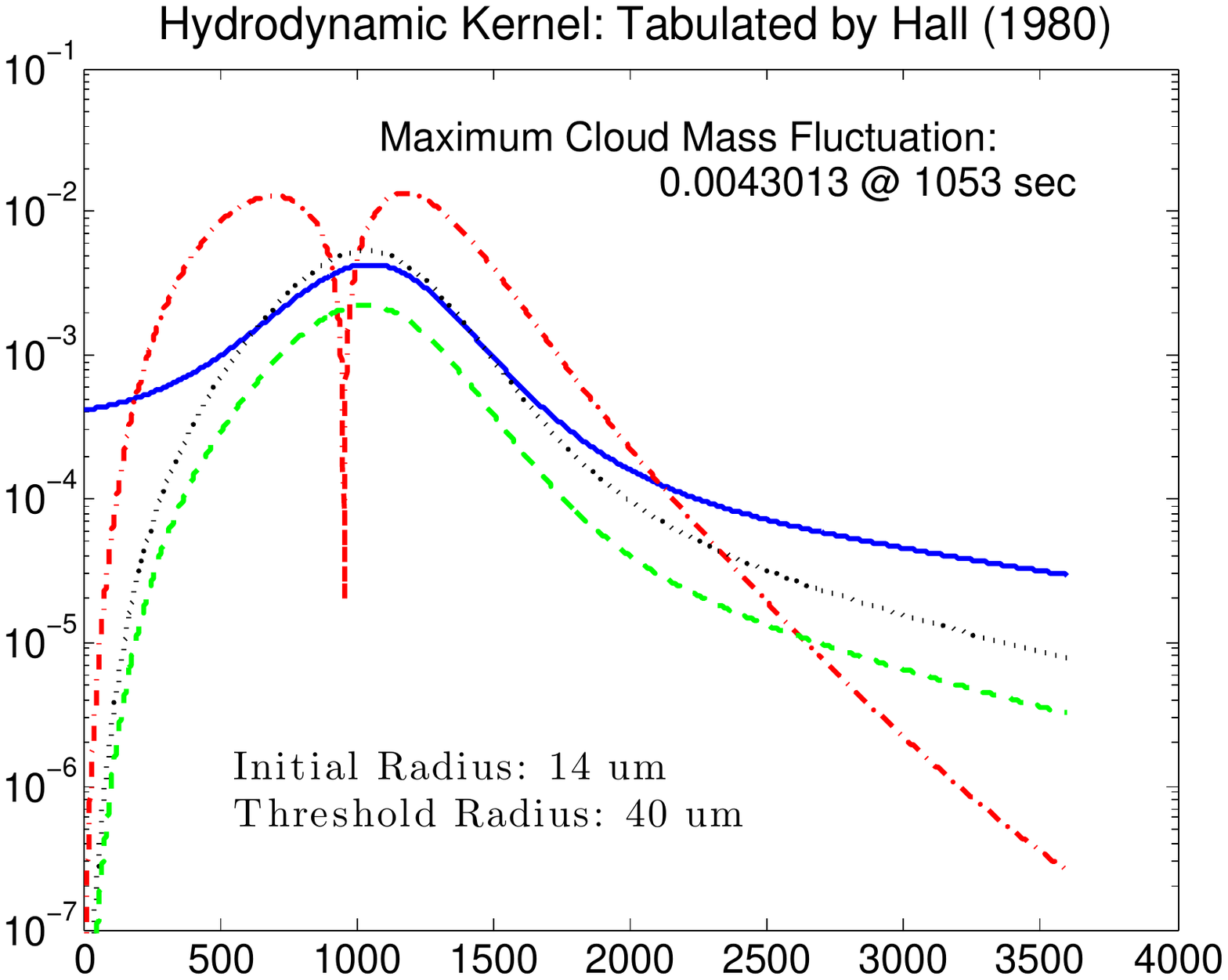}
	\includegraphics[width=.45\textwidth,trim=30 185 35 175,clip]{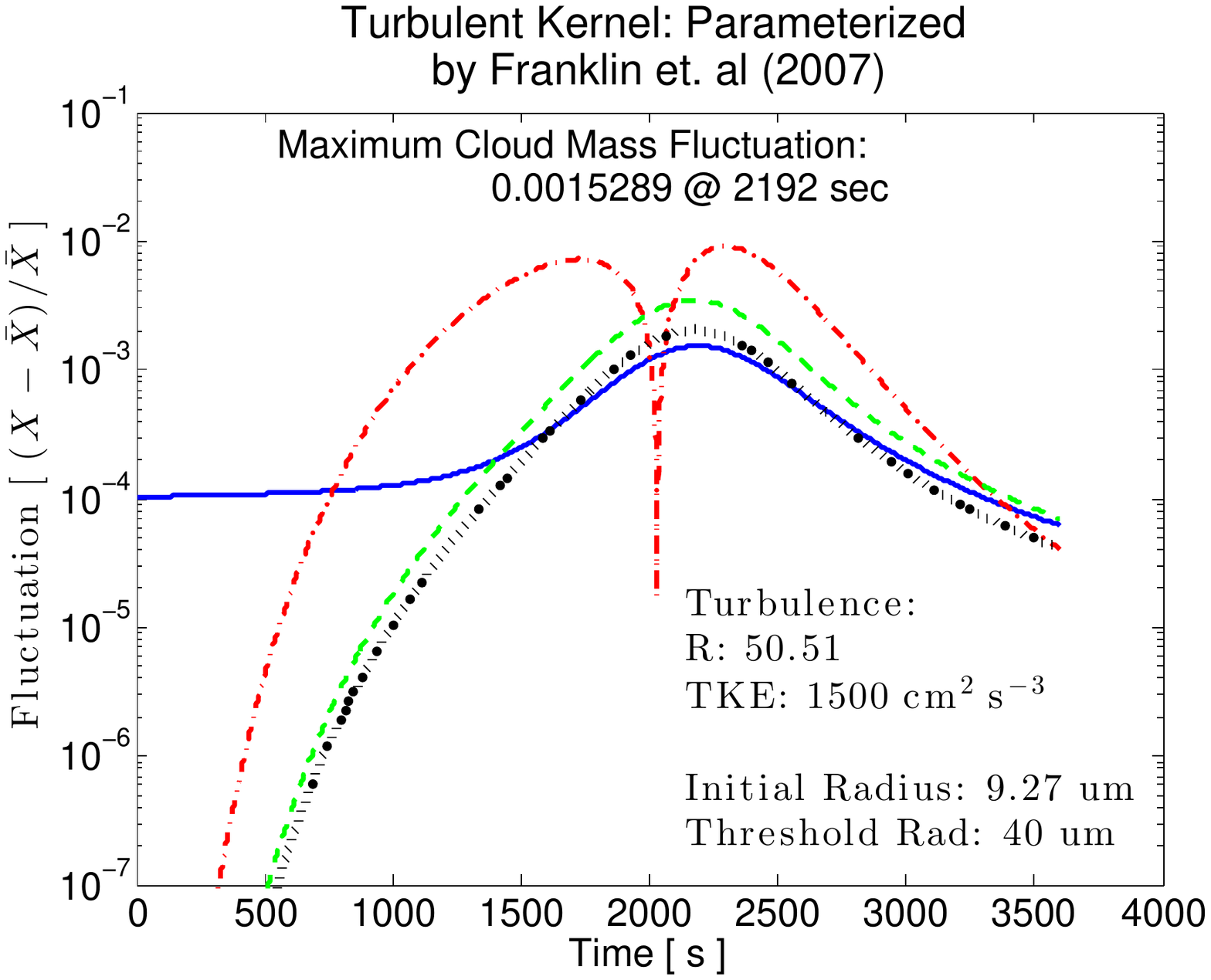}
	\includegraphics[width=.45\textwidth,trim=30 185 35 175,clip]{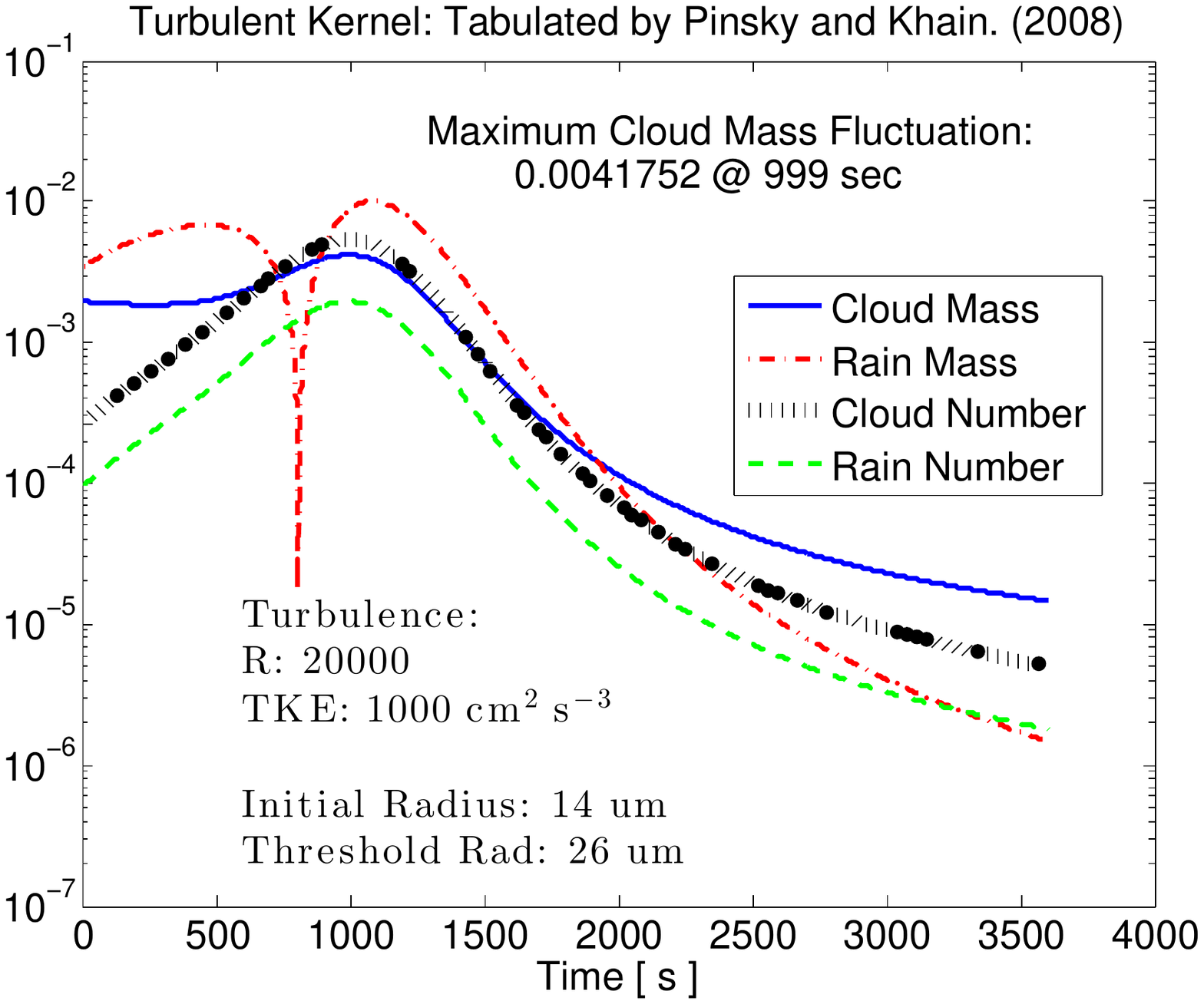} }
	\caption{Ratio of the fluctuations to the mean value for each of the four evolved quantities: $\delta{N_c}$/$ \ol{N_c}$, ~ $ \delta{N_r}/\ol{N_r}$, ~ $\delta{q_c}/\ol{q_c}$, ~ $\delta{q_r}/\ol{q_r}$.  Assuming ergodicity, the value of each evolved quantity was averaged over a 60 minute simulation.  The rain mixing ratio does not appear in the parameterization.\label{fig:DisregardFluctuationTerms} }   
\end{figure}

The graphical comparison of the fluctuations of the four evolving quantities to the respective temporal means of these quantities was done for four kernels:  (i) the piece-wise polynomial kernel used by Seifert and Beheng (2001), (ii) a hydrodynamic kernel with collection efficiencies tabulated by Hall (1980), (iii) a turbulent kernel parameterized by Franklin (2007), and (iv) a turbulent kernel with tabulated values by Pinsky and Khain (2008).  These results are shown in the four graphs in Figure \ref{fig:DisregardFluctuationTerms}.  

As shown in the four graphs in Figure \ref{fig:DisregardFluctuationTerms}, each of the evolving quantities, except for mixing ratio, has a maximum temporal fluctuation that is three orders of magnitude less than the quantities themselves.  Consequently, we can alternately take the approach of disregarding the cross terms containing, for example, $\ol{\delta{N_c}\delta{N_c}}$, thus providing a mean two-moment stochastic bulk rate parameterization shown in Equation \ref{eq:9parameters}.  The four mean stochastic bulk rate equations contain nine parameters which are the means of (i) temporal fluctuations and (ii) temporal product fluctuations of number concentrations and mixing ratios in regions on the 2-D domain in Figure \ref{fig:mass_aggregates}.   

These spectral fluctuations arise from the approximation of number concentration density and of mixing ratio density as the sums of means and fluctuations according to Equations ($\ref{eq:NC}$ and $\ref{eq:qc}$).  These parameters are present in terms which represent collision processes: cloud self-collection, rain self-collection, autoconversion, and accretion.  
\begin{equation}
\begin{aligned} \label{eq:9parameters}
\dif{\ol{N_c}}{t} = & -\left(\frac34-\frac{ 2\mu_{1nf}+ \mu_{1np}}2 \right)K_{cc}\ol{N_c^2}-K_{cr}\ol{N_cN_r}   \\
\dif{\ol{N_r}}{t} = & ~\left(\frac{1}4 + \frac{2\mu_{2nf}+\mu_{2np}}{2} \right) K_{cc}\ol{N_c^2}  -  \left(\frac34- \mu_{4np} \right) K_{rr}\ol{N_r^2}   \\
\dif{\ol{q_c}}{t} = &  -\left(\frac12- \left( \mu_{1nf}+ \mu_{1mf}+ \mu_{1mp} \right) \right) K_{cc}\ol{q_cN_c} - K_{cr}\ol{q_cN_r} \\
\dif{\ol{q_r}}{t} = & ~ \left(\frac12+ \mu_{2nf} + \mu_{2mf} + \mu_{2mp} \right) K_{cc} \ol{q_cN_c}  +  K_{cr}\ol{q_cN_r}.
\end{aligned}
\end{equation}
The omission of $q_r$ in Equation \ref{eq:9parameters} shows that the aggregate number of rain droplets affects accretion, but the aggregate rain mixing ratio does not.  This is in contrast to a model specifying individual droplet collisions whereby the mass of an individual rain droplet does affect the collection of a single cloud droplet \cite{HP97}.

\subsection{Reduced Parameter Set}  \label{sec:RPS}
Each of the four differential equations in \ref{eq:9parameters},  contains linear combinations of stochastic processes statistics.  In four instances, stochastic process within a linear combination describes the statistics of a single collision and coalescence process.  The means of fluctuations and of product fluctuations in a single region model the same phenomena and are linearly related in a single quantity.  Since we are simply collecting values for these parameters from data and are not examining the effects of modifying one parameter relative to another parameter, combining parameters in this linear manner does not pose a problem.  
\begin{equation}
\begin{aligned} \label{eq:LinearCombination}
\mu_{1n} = & 2\mu_{1nf}+ \mu_{1np} ~  & \text{ fluctuation of number concentration in } ~ \Omega_1   \\ 
\mu_{2n} = & 2\mu_{2nf}+\mu_{2np} ~&  \text{ fluctuation of number concentration in } ~ \Omega_2   \\
\mu_{1m} = &  \mu_{1nf} + \mu_{1mf} + \mu_{1mp}  ~ &  \text{ fluctuation of mixing in ratio in } ~ \Omega_1   \\
\mu_{1m} = &  \mu_{2nf} + \mu_{2mf} + \mu_{2mp} ~ & \text{ fluctuation of mixing in ratio in } ~ \Omega_2   \\ 
\end{aligned}
\end{equation}
The stochastic bulk rate parameterization thus reduces to
\begin{equation}
\begin{aligned}  \label{eq:condensedSBRP}
\dif{{N_c}}{t} = & -\frac14\left(3-2 \mu_{1n} \right)K_{cc}{N_c^2}-K_{cr}{N_cN_r}   \\
\dif{{N_r}}{t} = & ~  \frac{1}4 \left(1 + 2 \mu_{2n} \right) K_{cc}{N_c^2}   -  \left(\frac34- \mu_{4nc} \right) K_{rr}{N_r^2}   \\
\dif{{q_c}}{t} = &  -\left(\frac12-  \mu_{1m}  \right) K_{cc}{q_cN_c} - K_{cr}{q_cN_r} \\
\dif{q_r}{t} = & ~ \left(\frac12+ \mu_{2m} \right) K_{cc} {q_cN_c}  + K_{cr}{q_cN_r}.   \\
\end{aligned}
\end{equation}
The bars in Equation ($\ref{eq:condensedSBRP}$) are dropped, but the meaning of the two-moment evolved quantities will remain unchanged.

\subsection{Consistency of Number and Conservation of Mass} \label{sec:connumber_conmass}
The first term on the right-hand-side of the cloud number (first) equation in ($\ref{eq:condensedSBRP}$) contains the effects of both self-collection and auto-conversion.  Equation $\ref{eq:Ncee}$ distinguishes these effects, and separates the loss of droplets due to cloud self-collection from the gain of cloud droplets due to self-collection (i.e. the formation of a larger cloud droplet from two smaller could droplets).  The first three terms on the right-hand-side of Equation $\ref{eq:Ncee}$ represent the increase growth of cloud droplets due to self-collection.  The fourth term ($-K_{cc}N_c^2$) represents the total loss (not net loss) of cloud droplets due to both processes.  The coefficient of the first $K_{cc}N_c^2$ term in Equation \ref{eq:condensedSBRP} is rewritten to separate cloud droplet loss (cloud self-collection and auto conversion) from cloud droplet gain (only cloud self-collection): $-1/4(3-2\mu_{1n})=[1/4+2\mu_{1n}]-1$.  

Since two cloud droplets necessarily form either another cloud droplet or a new rain droplet, and we restrict collisions to binary ones, the following equation produces a constraint on the relationship between $\mu_{1n}$ and $\mu_{2n}$.  We equate the loss of cloud droplets to the gain of droplets resultant from that loss:
 \begin{equation} \label{eq:NumberCompatiblity}
 \underbrace{2~\left(\frac14\left(1+2 \mu_{1n} \right) \right) }_{\text{gain from cloud self-collection}} +   \underbrace{2~\left(\frac{1}4 \left(1 + 2 \mu_{2n} \right)  \right)}_{\text{gain from auto conversion}} -   \underbrace{1}_{\text{loss}}~=~0
 \end{equation}
Equation ($\ref{eq:NumberCompatiblity}$) reduces to $\mu_{1n} = -\mu_{2n} $.

Conservation of mass is addressed by noting the similarity of the integration of the accretion terms (which span a complete aggregate), with the integration of the regions $\Omega_1$ and $\Omega_2$ when they are combined.  The resultant integral also spans a complete aggregate and can be solved analytically.  Particularly, we have $n(z)n(y)K_{cc}$ integrated over the combined region yielding $K_{cc}N_c^2$.  The first loss term from Equation \ref{eq:Ncsecond} integrates to $-K_{cc}N_c^2$.  Combining the mass gain and mass loss terms we have: 
\begin{equation} \label{eq:conservationofmass}
K_{cc}N_c^2-K_{cc}N_c^2=0.
\end{equation}
Separating the integrals for mixing ratio over domains $\Omega_1$ and $\Omega_2$, the two resultant integrals do not span a complete aggregate and we must employ stochastic processes to close the equations as was detailed in Sections (\ref{sec:CM} and \ref{sec:RM}).  One stochastic term, $\mu_{1m}$, arises from the linear combination of the second, third, and fourth terms on the right hand side of Equation \ref{eq:qcee}.  The other stochastic term, $\mu_{2m}$, arises from the linear combination of the second, third, and fourth terms on the right hand side of Equation \ref{eq:qrev}.  The deterministic term is the fifth term on the right hand side of Equation \ref{eq:qcee}.  The coefficients of these three mixing ratio terms, which contain $K_{cc}N_c^2$, are set equal to the right hand side of Equation \ref{eq:conservationofmass}:
\begin{equation} \label{eq:conservationofmasstwo}
\left(\frac{1}2 + \mu_{1m} \right) + \left(\frac12+ \mu_{2m} \right) - 1=0.   \nonumber
\end{equation}
From which follows: $\mu_{1m} = -\mu_{2m}$.  Rewriting the four stochastic bulk rate equations using three independent parameters yields
\begin{equation}
\begin{aligned}  \label{eq:threedegreesSBRP}
\dif{{N_c}}{t} = & -\frac14\left(3-2 \mu_{1n} \right)K_{cc}{N_c^2}-K_{cr}{N_cN_r}   \\
\dif{{N_r}}{t} = & ~  \frac{1}4 \left(1 - 2 \mu_{1n} \right) K_{cc}{N_c^2}   -  \left(\frac34- \mu_{4nc} \right) K_{rr}{N_r^2}   \\
\dif{{q_c}}{t} = &  -\left(\frac12-  \mu_{1m}  \right) K_{cc}{q_cN_c} - K_{cr}{q_cN_r} \\
\dif{q_r}{t} = & ~ \left(\frac12- \mu_{1m} \right) K_{cc} {q_cN_c}  + K_{cr}{q_cN_r}  \\  
\end{aligned}
\end{equation}  
Conservation of mass for auto conversion follows intrinsically from the derivations and is clearly shown by the similarity of the first terms on the right hand sides of the mixing ratio equations in Equation \ref{eq:threedegreesSBRP}.  Conservation of mass for accretion follows from the $K_{cr}{q_cN_r} $ terms in these same two equations.

The physical meanings and effects of these final three parameters are as follows.
\begin{align}
\mu_{1n}   \hspace{1.0cm}  & \text{ controls the strength of cloud self-collection relative to auto-conversion }   \nonumber \\ 
\mu_{1m}   \hspace{1.0cm} &  \text{ controls the strength of auto-conversion }  \nonumber \\  
\mu_{4nc}  \hspace{1.0cm} & \text{ controls the strength of rain self-collection }  \nonumber \\   \nonumber
\end{align}  
Bounds on the values of these stochastic parameters will be established in Part II of this series of papers.

\section{Conclusion} \label{sec:Conclu}
Assumptions and simplifications are necessary to produce computationally affordable parameterizations that represent cloud microphysical processes.  Parameterizations of cloud microphysical processes over the past forty-five years have made assumptions regarding the droplet size distribution \cite{AS01,YL04,YL06}.  Many parameterizations depend on ad-hoc parameters \cite{MK00,AS01,YL04,YL06,CF08}.  Here, we presented a stochastic two-moment bulk parameterization of collision and coalescence which does not rely on any distribution of the droplet size spectrum, but rather assumes that there is a distribution and it has a mean.  All of the parameters in the stochastic parameterization have physical meaning, and their values can be recovered from data.  The parameters represent uncertainties in the first and second moments of time series of aggregate fluctuations of mixing ratio and number over defined portions of the droplet size spectrum.  

Other parameterizations have been restricted to a specific kernel \cite{AS01,YL04,YL06,CF08}.   The kernel in Franklin's parameterization was dependent on the turbulent strength, but limited in the range of turbulent kinetic energy used in that kernel.  The stochastic bulk parameterization can be used with a variety of kernels without any further derivations.  The value of the kernel at the mean cloud radius and mean rain radius is used in the stochastic bulk parameterizations.  Seifert and Beheng's parameterization used a kernel that contained both cloud and rain mean radii, and this analytic kernel could be directly applied to the resultant stochastic derivation.  Khairoutdinov and Kogan (2000) calibrated their parameters with data considered to be non-turbulent.  Thus their bulk parameterization is representative of one that employs a hydrodynamic kernel.  The second paper in this series applies the analytic kernel used by Seifert and Beheng to the new stochastic parameterization and precisely reproduces their parameterization while retaining the flexibility via the stochastic parameters to more accurately reproduce results from a detailed bin-based algorithm.

We assumed that any droplet distribution has a mean, and that the density can be constructed as the sum of the mean and fluctuations from the mean.  The flexibility of using any collision kernel in the stochastic parameterization comes at the expense of retaining only the zeroth order term in the 2D Taylor expansion of the collision kernel centred at either the cloud, or rain, mean mass.  This flexibility requires that the value of the collision kernel be computed, or retrieved from a look-up table, while the stochastic parameterization is being utilized by a climate model.  The stochastic bulk parameterization of cloud microphysical processes contains only three stochastic parameters, one for each three collision processes:  cloud-self collection, rain self-collection, and auto conversion.  In Part II, the stochastic parameterization in Equation \ref{eq:threedegreesSBRP} is validated against detailed simulations of the KCE (Eq. \ref{eq:KCE}) for the piece-wise polynomial kernel while the results from the hydrodynamic kernel and two turbulent kernels will be reported in future publications. \\

\textbf{Acknowledgements}
This research is part of D. Collins's Ph.D. thesis.  The research of B. Khouider is partly supported by a grant from the Natural Sciences and Engineering Research Council of Canada.  D. Collins' fellowship is partly funded through this grant.

\appendix

\numberwithin{equation}{section}
\section{Appendix: Compatibility Condition} \label{sec:CompatibilityCondition}
For any given droplet size of mass $x$, the number concentration density and the mixing ratio density at time $t$ are $n(x,t)$ and $q(x,t)$, respectively; and are related physically by
\be
q(x,t) = x n(x,t). \nonumber
\ee
Using the second relation in Equation $\ref{eq:PM}$ and this physical relation between the two densities, another expression for aggregate cloud mixing ratio is $q_c(t)=\int_0^{x^*} q(x,t) dx$.  The two compatibility conditions, one for the cloud partition of the spectrum and one for the rain partition, are
\begin{equation}
\begin{aligned}  \label{eq:CompatibilityCondition}
& \int_0^{x^*} xn(x,t) dx=\int_0^{x^*} q(x,t) dx, ~~~ \text{ and } ~~~  \\ & \lim_{x_m \rightarrow \infty} \int_{x^*}^{x_m} xn(x,t) dx=\lim_{x_m \rightarrow \infty}  \int_{x^*}^{x_m} q(x,t) dx. \nonumber
\end{aligned}
\end{equation} 

\textbf{Cloud Aggregate Compatibility} \\
The cloud aggregate approximation gives
\be
\int_0^{x^*} x \left( \frac{N_c(t)}{x^*}+\delta_c(\omega_1;x,t) \right) dx=\int_0^{x^*} \frac{q_c(t)}{x^*}+\gamma_c(\omega_2;x,t) dx.  \nonumber
\ee
Integrating the deterministic parts produces
\be
 \frac{x^*N_c(t)}{2}+\int_0^{x^*} x \delta_c(\omega_1;x,t) dx= q_c(t) +  \int_0^{x^*}\gamma_c(\omega_2;x,t) dx. \nonumber
\ee
The integral on the rhs is shown to be zero in Section $\ref{sec:SBPR}$, and the end of the proof is given in Equation $\ref{eq:mean0}$.  A slight rearrangement gives
\be \label{eq:CloudCompatibility}
\int_0^{x^*} x \delta_c(\omega_1;x,t) dx= q_c(t) - \frac{x^*N_c(t)}{2}= N_c \left( \ol x_c-\frac{x^*}2 \right)  
\ee
The relationship between the centre of mass of the cloud aggregate and the separation mass shows that the weighted number fluctuation within the cloud aggregate is negative when $2\ol x_c\le x^*$.  The negative values for the integral exist for most of the cloud evolution, and this can be verified with data.

\textbf{Rain Aggregate Compatibility}\\
The rain aggregate approximation gives
\be
\lim_{x_m \rightarrow \infty} \int_{x^*}^{x_m} x \left( \frac{N_r(t)}{x_m-x^*}+\delta_r(\omega_1;x,t) \right) dx= \lim_{x_m \rightarrow \infty} \int_{x^*}^{x_m} \frac{q_r(t)}{x_m-x^*}+\gamma_c(\omega_2;x,t) dx.  \nonumber
\ee
which produces the following expression for the weighted number fluctuation within the rain aggregate 
\be
\lim_{x_m \rightarrow \infty} \int_{x^*}^{x_m} x \delta_r(\omega_1;x,t) dx= q_r(t) - \lim_{x_m \rightarrow \infty} \frac{(x_m+x^*)N_r(t)}{2}.  \nonumber
\ee
Because droplets cannot grow to an infinite mass \cite{HP97}, take $x_m \gg x^*$ to be finite an have an expression for the  weighted number fluctuation within the rain aggregate that can be acquired numerically:
\be  \label{eq:RainCompatibility}
 \int_{x^*}^{x_m} x \delta_r(\omega_1;x,t) dx= q_r(t) - \frac{x_mN_r(t)}{2} = N_r \left( \ol x_r-\frac{x_m}2 \right).
\ee
The relationship between the centre of mass of the rain aggregate and the separation mass shows that the weighted number fluctuation within the rain aggregate is negative when $2\ol x_r\le x_m$.  The negative values for the integral exist for the entirety of the cloud evolution, and this can be also verified with data. 

In the current bulk cloud microphysics parameterization developed herein, the mass term which is added to the kinetic collection equation given by Equation \ref{eq:KCE} to get the first moment is absorbed by $q(x,t) = x n(x,t)$.   Therefore the lhs of Equations ($\ref{eq:CloudCompatibility}$ and $\ref{eq:RainCompatibility}$) do not appear in the derivations nor the final stochastic parameterization, and these compatibility conditions do not constrain the parameterization developed herein.

Any stochastic parameterization which uses the number concentration density approximation in Equation $\ref{eq:NC}$ and omits the mixing ratio density approximation in Equation $\ref{eq:qc}$ will contain the lhs of Equations ($\ref{eq:CloudCompatibility}$ and $\ref{eq:RainCompatibility}$)  in the derivations and the final parameterization.  Thus the utility of those parameterizations will be limited by conditions imposed by Equations ($\ref{eq:CloudCompatibility}$ and $\ref{eq:RainCompatibility}$).

\bibliographystyle{spmpsci}      
\bibliography{mybib_P1}   

\begin{thebibliography}{10}
\providecommand{\url}[1]{{#1}}
\providecommand{\urlprefix}{URL }
\expandafter\ifx\csname urlstyle\endcsname\relax
  \providecommand{\doi}[1]{DOI~\discretionary{}{}{}#1}\else
  \providecommand{\doi}{DOI~\discretionary{}{}{}\begingroup
  \urlstyle{rm}\Url}\fi

\bibitem{AB98}
Bott, A.: A flux method for the numerical solution of the stochastic collection
  equation.
\newblock \textit{J. Atmos. Sci.} \textbf{\textbf{55}}, 2284--2293 (1998)

\bibitem{BD12}
Devenish, B., Bartello, P.: Review article: Droplet growth in warm turbulent
  clouds.
\newblock J. Q. R. Meteorol. Soc. \textbf{138}, 1401--1421 (2012)

\bibitem{RD72}
Drake, R., Wright, T.: The scalar transport equation of coalescence theory: New
  families of exact solutions.
\newblock J. Atmos. Sci. \textbf{29}, 548--556 (1972)

\bibitem{CF08}
Franklin, C.: A warm rain microphysics parameterization that includes the
  effects of turbulence.
\newblock \textit{J. Atmos. Sci.} \textbf{\textbf{65}}, 1795--1816 (2008)

\bibitem{CF07}
Franklin, C., Vaillancourt, P., Yau, M.K.: Statistics and parameterizations of
  the effect of turbulence on the geometric collision kernel of cloud droplets.
\newblock \textit{J. Atmos. Sci.} \textbf{\textbf{64}}, 938--954 (2007)

\bibitem{WG13}
Grabowski, W.W., Wang, L.: Growth of cloud droplets in a turbulent environment.
\newblock \textit{Annu. Rev. Fluid Mech.} \textbf{\textbf{45}}, 293--324 (2013)

\bibitem{WH80}
Hall, W.D.: A detailed microphysical model within a two-dimensional dynamic
  framwork: Model description and preliminary results.
\newblock J. Atmos. Sci. \textbf{37}, 2486--2507 (1980)

\bibitem{MK00}
Khairoutdinov, M., Kogan, Y.: A new cloud physics parameterization in a large
  eddy simulation model of marine stratocumulus.
\newblock \textit{Mon. Wea. Rev.} \textbf{\textbf{128}}, 229--243 (2000)

\bibitem{SK93}
Krueger, S.: Linear eddy modeling of entrainment and mixing in stratus clouds.
\newblock J. Atmos. Sci. \textbf{50}(18), 3078--3090 (1993)

\bibitem{SK97}
Krueger, S., Chwen-Wei, McMurtry, P.A.: Modeling entrainment and finescale
  mixing in cumulus clouds.
\newblock J. Atmos. Sci. \textbf{54}(23), 2697--2712 (1997)

\bibitem{MvLW12}
Lier-Walqui, V., M., T.V., Posselt, D.: Quantification of cloud microphysical
  parameterization uncertainty using radar reflectivity.
\newblock Mon. Weather Rev. \textbf{140}, 3442--3466 (2012)

\bibitem{YL04}
Liu, Y., Daum, P.: Parameterization of the autoconversation process. {Part I}:
  Analytical formulation of the kessler-type parameterizations.
\newblock \textit{J. Atmos. Sci.} \textbf{\textbf{61}}, 1539--1548 (2004)

\bibitem{YL06}
Liu, Y., Daum, P., McGraw, R., Wood, R.: Parameterization of the
  autoconversation process. {Part II}: Generalization of the sundqvist-type
  parameterizations.
\newblock \textit{J. Atmos. Sci.} \textbf{\textbf{63}}, 1103--1109 (2006)

\bibitem{MP08}
Pinsky, M., Khain, A., Krugliak, H.: Collisions of cloud droplets in a
  turbulent flow. {Part V}: Applications of detailed tables of turbulent
  collision rate enhancment to simulation of droplet spectra evolution.
\newblock J. Atmos. Sci. \textbf{65}, 357--374 (2008)

\bibitem{DP10}
Posselt, D.J., Vukicevic, T.: Robust characterization of model physics
  uncertainty for simulations of deep moist convection.
\newblock Mon. Weather Rev. \textbf{138}, 1513--1535 (2010)

\bibitem{HP97}
Pruppacher, H., Klett, J.: Microphysics of Clouds and Precipitation.
\newblock \textit{Kluwer Academic Publishers}, Boston (1997)

\bibitem{AS01}
Seifert, A., Beheng, K.D.: A double moment parameterization for simulating
  autoconversion, accretion, and self-collection.
\newblock \textit{J. Atmos. Sci.} \textbf{\textbf{59-60}}, 265--281 (2001)

\bibitem{RS03}
Shaw, R.: Particle turbulent interactions in atmospheric clouds.
\newblock \textit{Annu. Rev. Fluid Mech.} \textbf{\textbf{35}}, 183--227 (2003)

\bibitem{MS02}
Simmel, M., Trautmann, T., Tetzlaff, G.: Numerical solution of the stochastic
  collection equation - comparison of the linear discrete method with other
  methods.
\newblock Atmos. Res. \textbf{61}, 135--148 (2002)

\bibitem{CS98}
Su, C.W., Krueger, S., McMurtry, P.A., Austin, P.: Linear eddy modeling of
  droplet spectral evolution during entrainment and mixing in cumulus clouds.
\newblock Atmos. Res. \textbf{47-48}(41-58) (1998)

\bibitem{LW07}
Wang, L., Xue, Y., Grabowski, W.: A bin integral method for solving the kinetic
  collection equation.
\newblock \textit{J. Comp. Physics} \textbf{\textbf{226}}, 59--88 (2007)

\bibitem{RW05b}
Wood, R.: Drizzle in stratiform boundary layer clouds. {Part II}: Microphysical
  aspects.
\newblock J. Atmos. Sci. \textbf{62}, 3034--3050 (2005)

\bibitem{RW02}
Wood, R., Field, P., Cotton, W.R.: Autoconversion rate bias in stratiform
  boundary layer cloud parameterizations.
\newblock Atmos. Res. \textbf{65}, 109--128 (2002)

\bibitem{IZ94}
Zawadzki, I., Fabry, F.: The development of drop size distribution is light
  rain.
\newblock J. Atmos. Sci. \textbf{51}(1100-1114) (1994)

\end{thebibliography}

\end{document}